\documentclass[aip,apl,superscriptaddress,reprint]{revtex4-1}
\usepackage{lineno,xcolor}
\usepackage{graphicx}
\newcommand{\enlarge}{\scalebox{1}}

\begin{document}

\title{Zero and Finite Energy Dirac Points in the Energy Band Structure of Magnetic Bilayer Graphene Superlattices}

\author{ C. Huy Pham$^{1,2}$, T. Thuong Nguyen$^{1,2}$, and V. Lien Nguyen$^{1,3,}$}
\email{nvlien@iop.vast.ac.vn}
\affiliation{ $^1$Theoretical and Computational Physics Department, Institute of Physics, VAST,  \\
10 Dao Tan, Ba Dinh Distr.,  Hanoi 10000,  Vietnam \\
$^2$ \enlarge{SISSA/International School for Advanced Study, Via Bonomea 265, I-34136 Trieste, Italy} \\
$^3$ \enlarge{Institute for Bio-Medical Physics, 109A Pasteur, $1^{st}$ Distr., Hochiminh City, Vietnam} }

\setlength{\textfloatsep}{1pt}
\begin{abstract}
Energy band structure of the bilayer graphene superlattices with $\delta$-function magnetic barriers and zero average magnetic flux is studied within the four-band continuum model, using the transfer matrix method. The periodic magnetic potential effects on the zero-energy Dirac point of pristine bilayer graphene are exactly analyzed. Magnetic potential is shown also to generate finite-energy Dirac points at the edges of Brillouin zone the positions of which and the related dispersions are determined in the case of symmetric potentials.  
\end{abstract}

\maketitle
\enlargethispage{2\baselineskip}
As is well-known from semiconductor physics, the energy band structure of materials could be essentially modified by an external periodic potential, resulting in unusual transport and optical properties. That is why the energy band structure of graphene under a periodic potential (graphene superlattice) was extensively studied from the early days of graphene physics. For single layer graphene superlattices (SLGSLs), the energy band structure has been in detail examined in a number of works for periodic potentials of different natures (electric \cite{park,brey,barbier,huy} or magnetic \cite{ghosh,masir,snyman,dellan,lequi}) and different shapes (Kronig-Penny \cite{park,huy,masir,lequi}, cosine \cite{brey} or square \cite{barbier}). Interesting findings have been reported, such as a strongly anisotropic renormalization of the carrier group velocity and an emergence of extra Dirac points (DPs) in the band structure of electric SLGSLs \cite{park,brey,barbier,huy} or an emergence of finite-energy DPs in the band structure of magnetic ones \cite{snyman,dellan,lequi}. There are fewer works concerning the energy band structure of bilayer graphene superlattices (BLGSLs) and they are all devoted to the case of electric potentials \cite{peeters,killi,tan}. The most impressive features observed in the band structure of  the electric BLGSLs studied (with different potential shapes: $\delta$-function \cite{peeters}, rectangular \cite{killi}, or sine \cite{tan}) are an emergence of a pair of new zero-energy DPs or an opening of a direct band gap, depending on the potential strength. These unusual features have been observed in the band structure of only electric BLGSLs and, moreover, they are assumed common for all electric BLGSLs with any potential shape, providing the average potential to be zero. Then, a question should be certainly raised of whether periodic magnetic potentials could bring about similar features in the energy band structure of bilayer graphene (BLG). Unfortunately, to our knowledge, no works on the band structure of BLGSLs with magnetic potentials have been reported. Note that while sharing with single layer graphene many properties important for electronics applications, BLG exhibits the privileges, including the ability to open a band gap in the energy spectrum and to turn it flexibly by an external electric field \cite{castro,cann}. On the other hand, the effects of a magnetic potential and those of an electric one on the energy band structure of graphene might be related to each other \cite{louie}.   

The purpose of the present work is to study the energy band structure of BLGSLs with a magnetic potential (magnetic BLGSLs - MBLGSLs) which are arisen from an infinitely flat Bernal-stacked BLG in a periodic magnetic field with zero average magnetic flux as schematically described in Fig.1$(a)$. The magnetic field is assumed to be uniform in the $y$-direction and staggered as $\delta$-function barriers of alternate signs, $\vec{B}_0$ and $- \vec{B}_0$, along the $x$-direction. The corresponding vector potential $A(x)$ can be then described as the standard Kronig-Penney potential with $d_B$ the barrier width, $d_W$ the well width, $d = d_B + d_W$ the superlattice period, and $A_0 = B_0 l_B$ the potential strength ($l_B = \sqrt{\hbar / eB_0}$ the magnetic length and $e$ the elementary charge) [Fig.1$(b)$]. For such the periodic magnetic potentials the only way of breaking the symmetry is associated with a difference between $d_B$ and $d_W$. So the parameter $q = d_W / d_B$ is introduced to describe the asymmetric effects and the MBLGSLs with $q = 1$ ($q \neq 1$) will be referred to as symmetric (asymmetric) MBLGSLs. Thus, within the model discussed the periodic magnetic potential is entirely characterized by the three parameters: $A_0$, $d$, and $q$. Such a $\delta$-function model will be hold as long as de Broglie wavelength of quasi-particles is much larger than the typical width of magnetic barriers \cite{ghosh}. Actually, this magnetic potential model is the same as that used for SLGSLs in Refs.\cite{ghosh,barbier,lequi}, but the structure studied here is BLG.

To justify the consideration, we will ignore intervalley scattering assuming the widths $d_{B(W)}$ are much larger than the lattice constant in graphene. All spin-related effects are also neglected. Besides, potentials on both graphene layers are assumed to be the same at a given $(x,y)$-point. Under these conditions the low-energy excitations near one original Dirac point (say, $K$) in the energy band structure can be generally described in the four-band continuum nearest-neighbor, tight-binding model with the Hamiltonian  
\begin{equation}
  H \ = \ \left(  \begin{array}{cccc}
        0  & \ v_F \hat{\pi}  &  \ t_\perp  &  \ 0  \\
        v_F \hat{\pi}^+  &  \ 0 &  \ 0  &  \ 0       \\
        t_\perp  &  \ 0  &  \  0  &  \ v_F \hat{\pi}^+  \\
        0  &  \ 0  &  \ v_F \hat{\pi}   &  \  0
    \end{array}   \right) \   ,
\end{equation}
where  $\hat{\pi} = p_x + i p_y$, $v_F = \sqrt{3} t a / (2 \hbar ) \approx 10^6 \ m/s$ is the Fermi velocity, $t \approx 3 \ eV$ is the intralayer nearest-neighbor hopping energy, $a = 2.46 \AA$ is the lattice constant of graphene, and $t_\perp \approx 0.39 \  eV$ is the interlayer nearest-neighbor hopping energy. The magnetic field effect is here accounted for by the momentum operator $\vec{p} = (p_x , p_y ) \equiv - i \hbar \vec{\nabla} + e \vec{A}$. The Hamiltonian of eq.(1) is limited to the case of symmetric on-site energies \cite{cann}. 

\begin{figure} [t]
\begin{center}
\includegraphics[trim = 0in 1in 0in 0in,width=9.0cm,height=4.0cm]{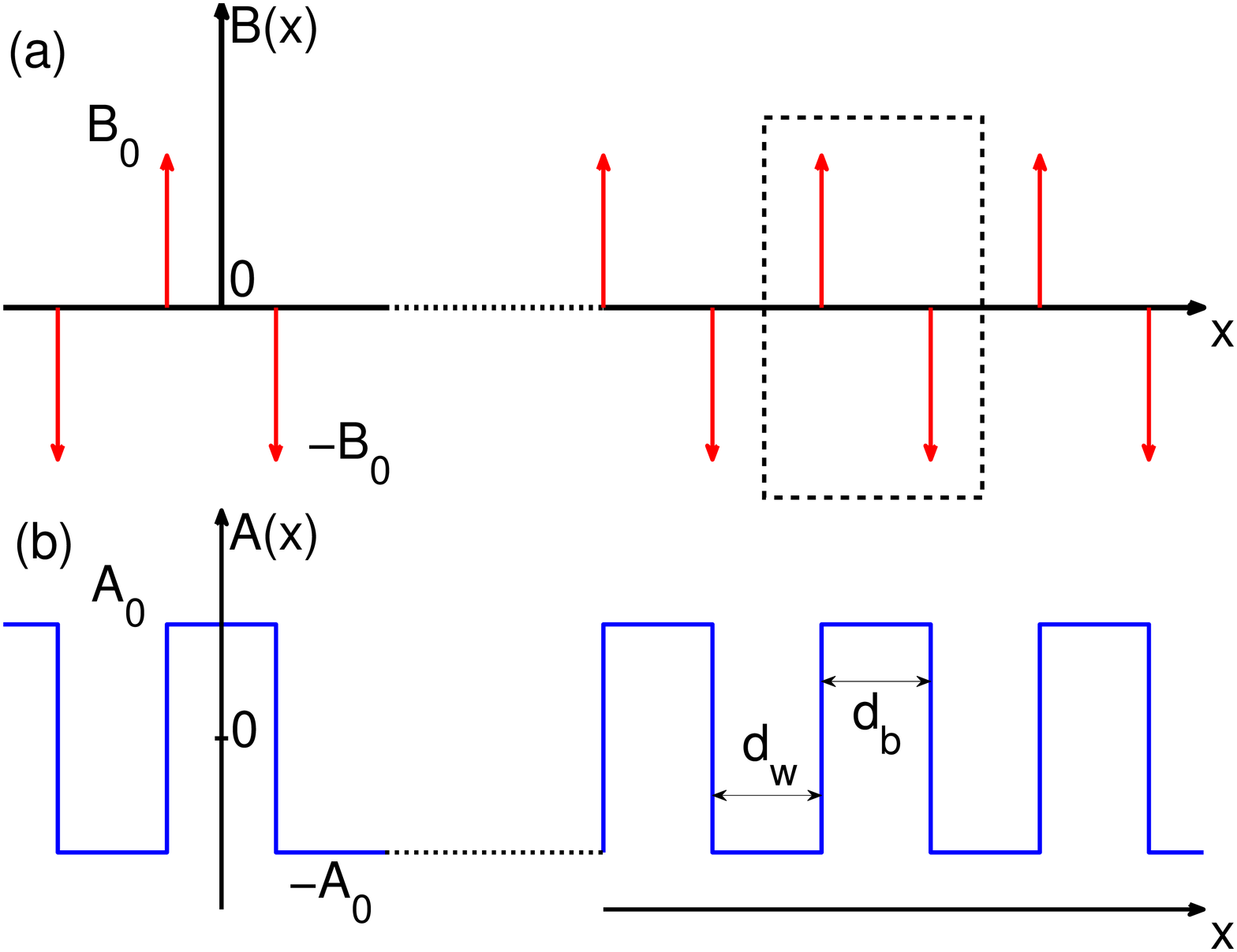}
\caption{
(color online)  Model of MBLGSLs under study: $(a)$ Periodic $\delta$-function magnetic barriers of alternate signs, $\vec{B}_0$ and $- \vec{B}_0$ [red arrows] and $(b)$ Corresponding 1D periodic vector potential $\vec{A}(x)$ [blue curve] with $A_0$ the potential strength, $d_B$ the barrier width, and $d_W$ the well width [the period $d = d_B + d_W$]. The dashed-line box in $(a)$ describes the unit cell in $T$-matrix calculations.
}
\end{center}
\end{figure}
Due to a periodicity of the potential $A(x)$ the time-independent Schr\"{o}dinger equation $H \Psi = E \Psi$ for the Hamiltonian $H$ of eq.(1) could be most conveniently solved using the transfer matrix method \cite{peeters,chau}, which generally reduces the energy spectrum problem to solving the equation (see Supplementary Material \cite{suppl}):
\begin{equation}
det \ [ \ T \ - \ e^{i k_x d} R^{-1}_I (d) \ ] \ = \ 0 , 
\end{equation} 
where $k_x$ is the Bloch wave vector and $T$ and $R_I$ are matrices, depending on the Hamiltonian under study. In the case of SLGSLs, when the Hamiltonian $H$ and, therefore, $T$ and $R_I$ are $2 \times 2$ matrices, equation (2) can be analytically solved that gives straightaway a general expression for the dispersion relation $E(\vec{k})$  \cite{lequi}. For MBLGSLs in the four-band model of eq.(1), in general, equation (2) with $(4 \times 4)$-matrices $T$ and $R_I$ is too complicated to be solved analytically, so we have solved it numerically and show in Figs.2 and 3, for example, the structure of several most important minibands.

{\sl Zero-energy DP}.- To examine the zero-energy DPs, Fig.2 presents the lowest conduction and the highest valence minibands for the MBLGSLs with potentials of $A_0 = 0.5$ and $d = 4$ in three cases: $(a) \ q = 1$ (symmetric potential), $(c) \ q  = 1.5$, and $(e) \ q = 0.5$ (asymmetric potentials). The boxes $(b)$, $(d)$, and $(f)$ present the contour plots of the lowest conduction miniband for the energy spectra shown in $(a)$, $(c)$, and $(e)$, respectively. (Due to a symmetry of spectra with respect to the $(E = 0)$-plane the analysis is hereafter concentrated on the positive energy part). For comparison, we remind that in the energy band structure of the pristine BLG there is a single zero-energy DP located at $\vec{k} = 0$, in the vicinity of which the dispersion has the parabolic double cone shape: $E = \pm \hbar^2 k^2 / 2 m$ with the isotropic mass $m = t_\perp / 2 v_F^2$ \cite{cann}. Hereafter, for convenience, dimensionless quantities are introduced: energies in units of $t_\perp$, $x$ (or $d$) in $ (\hbar v_F / t_\perp )$, and $k_{x(y)}$ in $(t_\perp / \hbar v_F )$ with $t_\perp$ and $v_F$ given above.

\begin{figure} [t]
\begin{center}
\includegraphics[trim = 0in 0.5in 0in 0in,width=9.0cm,height=7.0cm]{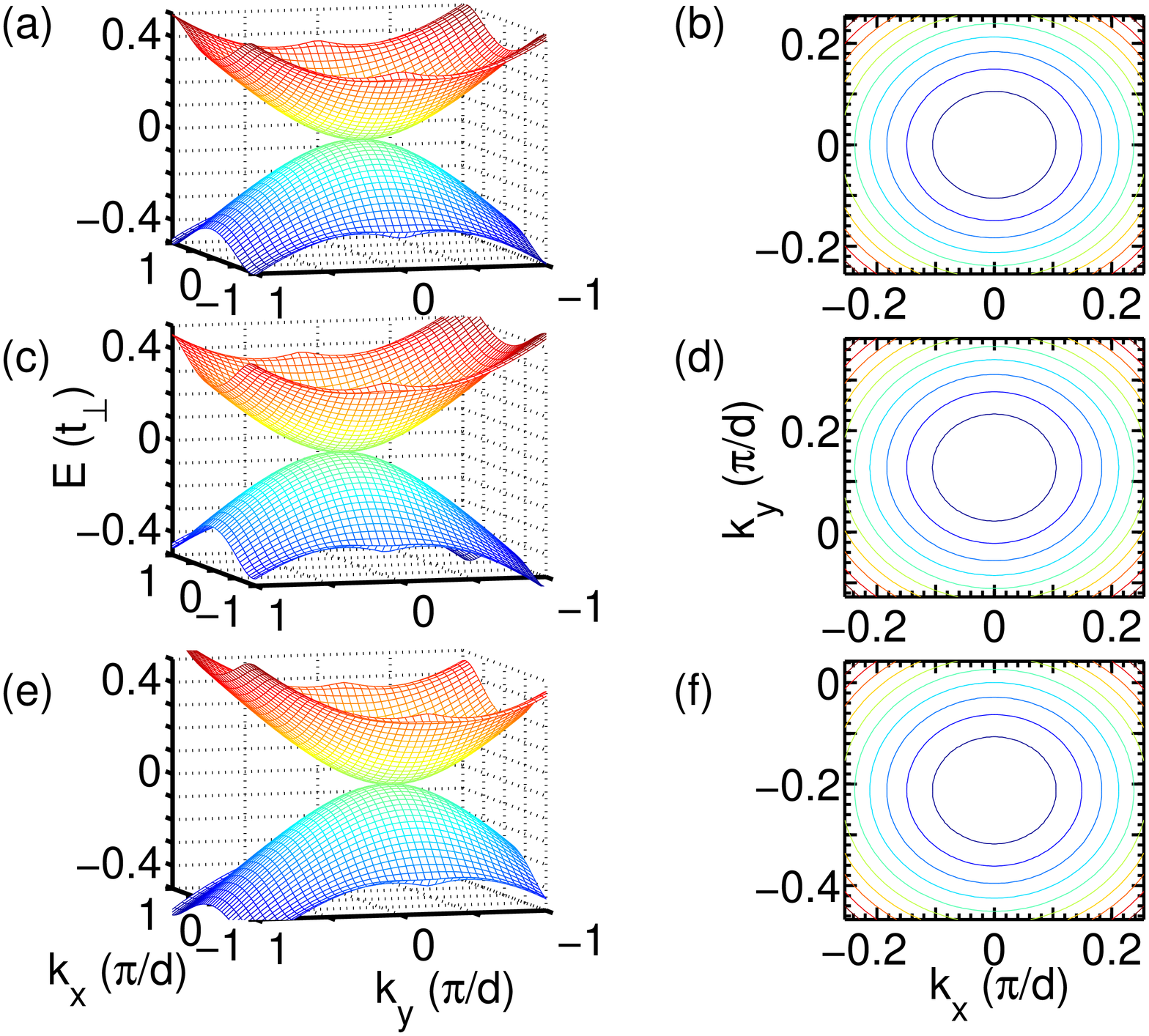}
\caption{
(color online) Zero-energy DP. Lowest conduction and highest valence minibands [$(a)$,$(c)$, and $(e)$] and corresponding contour plots [$(b)$ to $(a)$, $(d)$ to $(c)$, and $(f)$ to $(e)$] are shown in three cases: $q = 1$ [$(a,b)$], $q = 1.5$ [$(c,d)$], and $q = 0.5$ [$(e,f)$]. In all the cases: $d = 4$ and $A_0 = 0.5$. DP is located at $(E, k_x , k_y ) = (0, 0, 0)$ in $(a,b)$; $(0, 0, 0.1)$ in $(c,d)$; and $(0, 0, - 0.167)$ in $(e,f)$. All contour plots show isotropic dispersions.
}
\end{center}
\end{figure}
In the case of $q = 1$ (symmetric potential), Fig.2$(a)$ shows a clear zero-energy DP  at $(\vec{k} = 0)$ with an isotropic dispersion [see contour plot in $(b)$] like the original zero-energy DP for pristine BLG. In the case of $q \neq 1$, however, asymmetric magnetic potential moves the zero-energy DP along the $(k_x = 0)$-direction either in positive [if $q > 1$, see Figs.2$(c,d)$] or in negative direction of $k_y$ [if $q < 1$, see Figs.2$(e,f)$], while related dispersions are all still isotropic [see contour plots in $(d)$ and $(f)$]. Actually, the effects of periodic magnetic potential on the zero-energy DP sensitively depend on the potential parameters $A_0$, $d$, and $q$ and could be quantitatively understood from eq.(2) (see Supplementary Material \cite{suppl}). 

First, concerning the location of DP, we have arrived at the belief that the magnetic potential [along the $x$-direction] changes only the $k_y$-coordinate of the zero-energy DP, moving it from $(E = 0, k_x = 0, k_y = 0)$ [in the absence of potential] to $(E = 0, k_x = 0, k_y = k_y^{(a)}  = [(q - 1)/(q + 1)] A_0 )$ [in the presence of potential with $A_0$ and $q$]. Numerical solutions of eq.(2) well support this estimation for the location of zero-energy DP (see Fig.2, where the $(E, k_x )$-coordinates of all the DPs observed are $(0,  0)$, while their $k_y$-coordinates are depending on $q$:  $k_y = k_y^{(a)} = 0, \ 0.1$, and $\approx - 0.167$ in $(a,b) [q = 1]$, $(c,d) [q = 1.5]$, and $(e,f) [q = 0.5]$, respectively). Note that $k_y^{(a)}$ does not depend on the period $d$. 

Next, to check the dispersion relation associated with the zero-energy DP identified we expand eq.(2) to the lowest order in $E$, $k_y$, and $k_x$ in the vicinity of its location, $(E = 0, k_x = 0, k_y = k_y^{(a)})$, that leads to
\begin{equation}
E \ = \ \pm \frac{\hbar^2}{ 2 m^* } [ \ k_x^2 + (k_y - k_y^{(a)})^2 \ ] , 
\end{equation}
with the mass $m^*$ depending on $A_0$, $d$, and $q$:
\begin{equation}
m^* \ =  \ m  \ \frac{2A_0 d q}{ (q + 1)^2 \sinh (2A_0 d q / (q + 1)^2)}  .
\end{equation} 
\begin{figure} [t]
\begin{center}
\includegraphics[trim = 0in 0.5in 0in 0in,width=8.0cm,height=4.5cm]{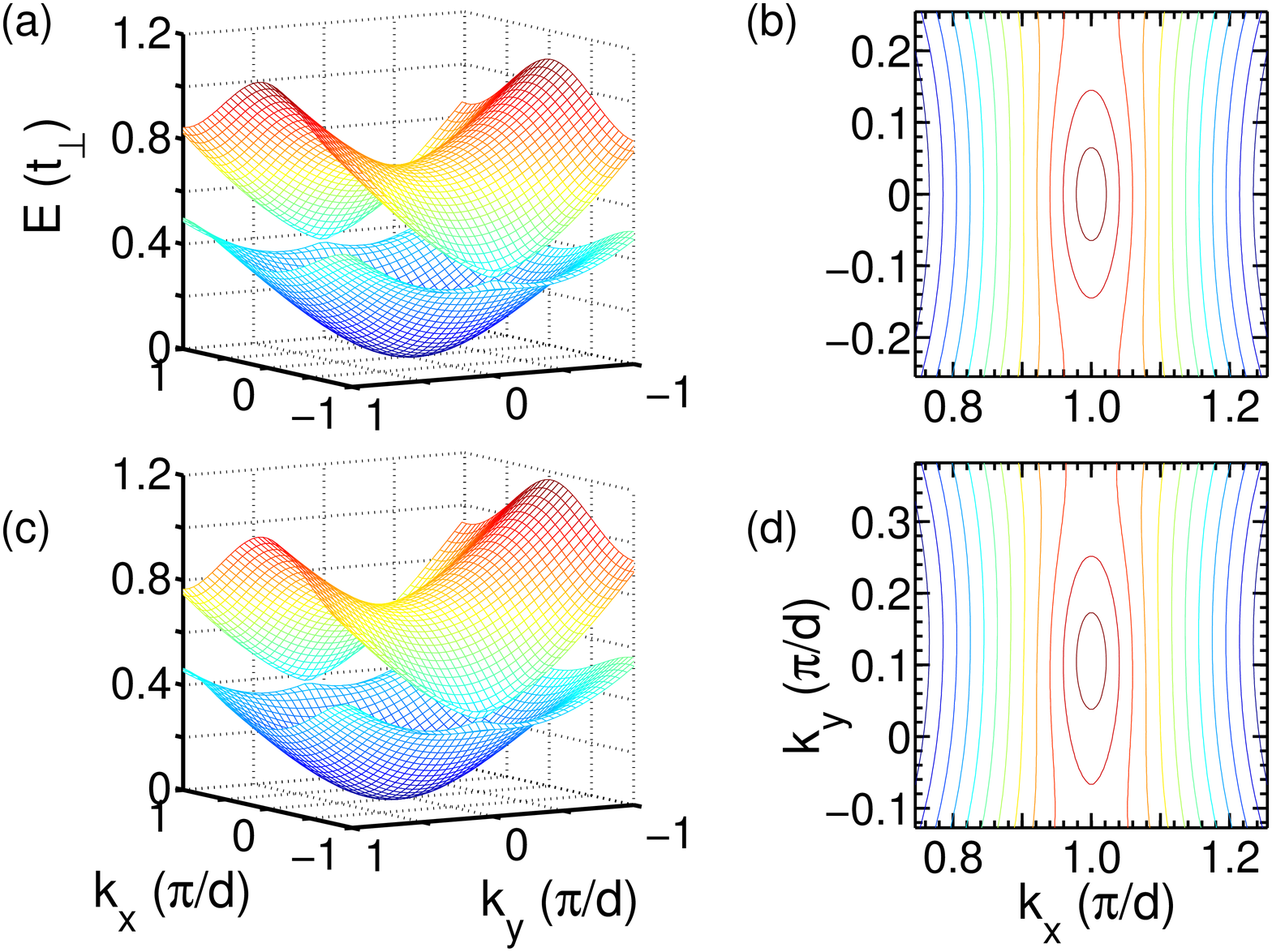}
\caption{
(color online) Finite-energy DPs. Two lowest conduction minibands [$(a)$ and $(c)$] and corresponding contour plots [$(b)$ to $(a)$ and $(d)$ to $(c)$] are shown in two cases: $q = 1$ [$(a,b)$] and $q = 1.5$ [$(c,d)$]. In both cases: $d = 4$ and $A_0 = 0.5$. Clealy, there is always a DP at the edge of Brilouin zone, $k_x = \pi /d$ [or, in equivalence, $k_x = - \pi /d$], which is located at $(E, k_y ) = (E_1 \approx 0.37 \ t_\perp , 0)$ in $(a)$ or $(E_1 \approx 0.375 \ t_\perp, k_y^{(f)} \approx 0.09)$ in $(c)$. Contour plots $(b)$ and $(d)$ show anisotropic dispersions.
}
\end{center}
\end{figure}
While the obtained relation of eq.(3) shows a parabolic dispersion with an isotropic double cone shape like that of original zero-energy DP for the pristine BLG, the isotropic mass $m^*$ is however depending on the potential parameters ($A_0 , d$ and $q$). It is clear from eq.(4) that $(i)$ $m^*$ is always less than $m$ ($= t_\perp / 2 v_F^2$ for the pristine BLG), $(ii)$ $m^* / m$ monotonously decreases with increasing $A_0$ and/or $d$, and $(iii)$ in the $m^* (q)$-dependence there is a single minimum at $q = 1$ (i.e. for symmetric potentials). Thus, the shift of location in the $k_y$-coordinate and the dependence of mass on the potential parameters are the two effects a periodic magnetic potential can induce on the zero-energy DP in the energy band structure of BLG. For comparison, remind again that the electric periodic potentials generate either a pair of new zero-energy DPs or a direct band gap \cite{peeters,killi,tan}.  

{\sl Finite-energy DPs}.- Fig.3 shows the two lowest conduction minibands [$(a)$ and $(c)$] and the corresponding contour plots [$(b)$ and $(d)$] for the same MBLGSLs as those studied in Figs.2$(a,b)$ and $(c,d)$, respectively. (Noting again the symmetry of the energy spectra with respect to the $(E = 0)$-plane). In both the cases, $q = 1$ $(a,b)$ and $q = 1.5$ $(c,d)$, the touching points are clearly seen at the edges of the Brillouin zone, $(k_x = \pm \pi / d, k_y = k_y^{(f)})$, where $k_y^{(f)}$ depends on $A_0$ and $q$ and equals to zero for symmetric potentials [Fig.3$(a,b)$]. Similar finite-energy touching points are also existed at higher energies (unshown). Describing quantitatively the touching points observed is generally beyond our ability except the case of symmetric potentials, when the locations of these points as well as related dispersions can be found in the same way as that used above for the zero-energy DP.   

\begin{figure} [t]
\begin{center}
\includegraphics[trim = 0in 1.0in 0in 0in,width=9.0cm,height=5.0cm]{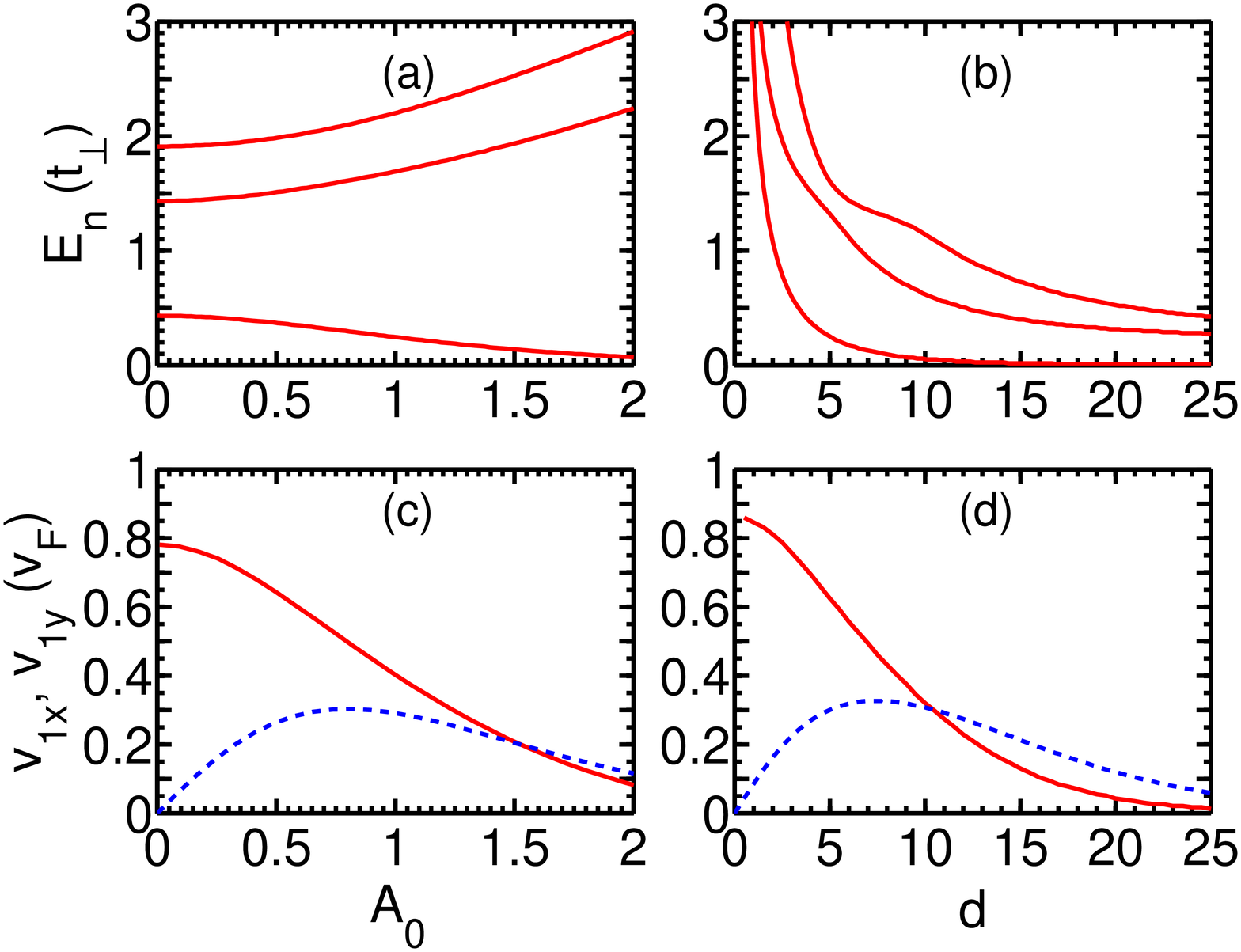}
\caption{
(color online) Three lowest from $E_n$ determined in eq.(5) are plotted versus $A_0$ [$(a)$ for $d = 4$] or $d$ [$(b)$ for $A_0 = 0.5$], $n = 1, 2$, and 3 (from bottom). In $(c)$ or $(d)$ velocities $v_{1x}$ [red solid line] and $v_{1y}$ [blue dashed line] in eq.(6) versus $A_0$ [$(c)$ for $d =  4$] or $d$ [$(d)$ for $A_0 = 0.5$], respectively.
}
\end{center}
\end{figure}
Indeed, in the case of symmetric potentials, $q = 1$, as can be seen in Figs.3$(a,b)$ [and can be directly checked using eq.(2)], the finite-energy touching points are located at ($k_x = \pm \pi / d,  k_y = 0$). Substituting these wave numbers into eq.(2), we obtain the relation:
\begin{eqnarray}
 &  &  \cos (k_1 d /2) \cos (k_2 d/2) -  \nonumber \\ 
 &  &  (A_0^2 / k_1 k_2 ) \sin (k_1 d/2) \sin (k_2 d/2) \ = \ 0 ,
\end{eqnarray}
where $k_{1(2)} = \sqrt{E^2 \pm E - A_0^2}$. Given $A_0$ and $d$, solving this equation we can fix the energy-coordinates $E_n$ of all finite-energy touching points of interest (which are all arranged in pairs). For example, in Figs.4$(a)$ or $(b)$ three lowest energies $E_n$ are plotted as a function of $A_0$ [for $d = 4$] or $d$ [for $A_0 = 0.5$], respectively. Except the lowest energy $E_1$ in Fig.4$(a)$ which shows an decrease with increasing $A_0$, all the energies $E_n$ calculated seem to increase with the potential strength $A_0$ [two higher curves in Fig.4$(a)$], but decreasing with increasing the potential period $d$ [Fig.4$(b)$]. Particularly, Figs.3$(a)$ shows $E_1 \approx 0.37 \ t_\perp$. Thus, eq.(5) gives energy-positions $E_n$ of all finite energy touching points which are located at $(k_x = \pm \pi / d, k_y = 0)$ in the case of symmetric magnetic potentials.

\begin{figure} [t]
\begin{center}
\includegraphics[trim = 0in 1.0in 0in 0in,width=9.0cm,height=5.0cm]{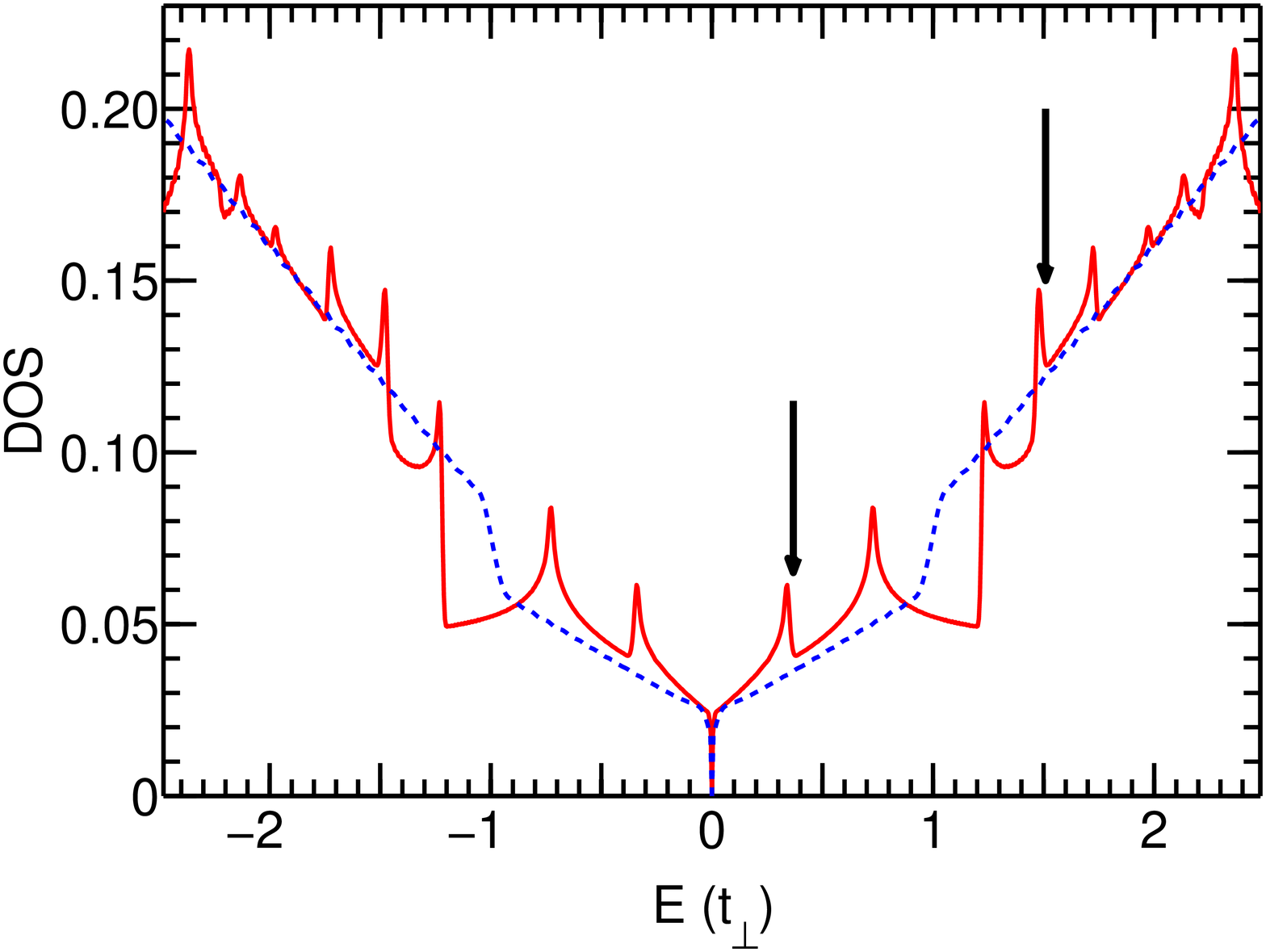}
\caption{
(color online) DOS for the MBLGSL with band structure presented in Fig.3$(a)$ [red solid line] and that for the pristine BLG [blue dashed line] are compared. Arrows indicate energy positions $E_1$ and $E_2$ of lowest finite energy DPs.
}
\end{center}
\end{figure}
Further, by expanding eq.(2) in the vicinity of the touching points fixed, $(E = E_n , k_x = \pi / d, k_y = 0)$, we arrive at the linear dispersion relation:
\begin{equation}
E - E_n \ = \ \pm  \sqrt{ v_{nx}^2 ( k_x - \pi /d )^2 + v_{ny}^2 k_y^2 } ,
\end{equation}
where $v_{nx}$ and $v_{ny}$ are carrier group velocity components depending on $A_0$ and $d$. Due to the double cone shape of these dispersions the finite-energy touching points fixed could be merely referred to as the finite-energy DPs. Unfortunately, we are unable to derive analytical expressions for $v_{nx}$ and $v_{ny}$, so for instance we show in Figs.4$(c)$ or $(d)$ the numerical values of $v_{1x}$ [solid line] and $v_{1y}$ [dashed line] for the lowest (and most important) finite-energy DP ($n = 1$) plotted against $A_0$ or $d$, respectively. At small $A_0$ and/or $d$  a large difference between the two velocities, $v_{1x} >> v_{1y}$, demonstrates a strongly anisotropic dispersion. Given $d$ (or $A_0$) there exists a single value of $A_0 = A_0^{(c)}$ (or $d = d^{(c)}$), where $v_{1x} = v_{1y}$ showing an isotropic dispersion [$A_0^{(c)} \approx 1.5 $ in Fig.4$(c)$ or $d^{(c)} \approx 10.435 $ in Fig.4$(d)$]. Beyond this point an anisotropy in dispersion is recovered, but it is much weaker than in the region of small $A_0$ and/or $d$. Returning to Figs.3$(a,b)$ with $A_0 = 0.5$ and $d = 4$, we have $v_{1x} / v_{1y} \approx 2.1 $. For higher finite-energy DPs ($n = 2, 3, ...)$ calculations show much more complicated $A_0$- and $d$-dependences of velocities (unshown) that demonstrate a strongly anisotropic dispersion at small as well  as large values of $A_0 / d$ except a single point where $v_{nx} = v_{ny}$. 

In the opposite case of asymmetric potentials, $q \neq 1$ [Figs.3$(c,d)$], we are able only to qualitatively comment that the finite-energy DPs with linear dispersion should be still generated at $k_x = \pm \pi / d$, but at non-zero $k_y = k_y^{(f)}$ and at energies which depend on potential parameters in the way much more complicated than eq.(5) [in Figs.3$(c,d)$ $k_y^{(f)} \approx 0.09$ and $E_1 \approx 0.375 \ t_\perp $]. Note that for BLGSLs with electric $\delta$-function potentials the finite-energy touching points are also generated at $(k_x = 0, k_y = 0)$, the related dispersions are however direction-dependent in the sense that they are linear or parabolic in the $k_x$- or $k_y$-direction, respectively \cite{huy2}.  

Finally, Fig.5 compares the density of state (DOS) for the band structure of the MBLGSL studied in Fig.3$(a)$ [red solid line] and that of the pristine BLG [blue dashed line]. Clearly, the periodic magnetic potential makes the DOS rather fluctuated (comparing to that of pristine BLG). The central minimum in the solid curve (at $E = 0$) is related to the zero-energy DP. The local dips at finite energy in this DOS is associated with the finite-energy DPs at $E_1$ and $E_2$ [indicated by arrows], whereas the peaks are located at the bending points between these DPs. Such a potential induced behavior of DOS should be manifested in transport properties of the structure.

In summary, we have studied the effects of a periodic $\delta$-function magnetic field with zero average flux (say, along the $x$-direction) on the energy band structure of BLGs. It was shown that the magnetic potential $(i)$ may move the location of the original zero-energy DP along the $(k_x = 0)$-direction to a finite $k_y$-coordinate, keeping the double cone dispersion isotropic and parabolic with the carrier mass which is however depending on the potential; and $(ii)$ generates the finite-energy DPs with a linear and anisotropic dispersion at the edges of the Brillouin zone. In the case of symmetric magnetic potentials the position and the dispersion are exactly determined for all the DPs of interest, while in the case of asymmetric potentials this can be done for the zero-energy DP only. We assume that these findings should robust against changes in the shape of the magnetic potential, providing the average flux to be zero. \\
{\sl Acknowledgments}.- This work was financially supported by Vietnam National Foundation for Science and Technology Development under Grant No. 103.02-2013.17.

\end{document}